\documentclass[showpacs, amsmath,amssymb, aps,showkeys]{revtex4}
\usepackage{amsmath,amssymb,amsfonts,latexsym,float,graphics,epsfig,bm}
\usepackage{graphicx}

\catcode`@=11 \@addtoreset{equation}{section}

\newcommand{\be}{\begin{equation}}
\newcommand{\ee}{\end{equation}}

\begin{document}
\title{Optical properties of null geodesics emerging from dynamical systems}
\affiliation{%
Department of Mathematics,\\  Khalifa University of Science and Technology,\\ Main Campus, Abu Dhabi,\\ United Arab Emirates}
\author{D. Batic}
\email{davide.batic@ku.ac.ae}
\affiliation{%
Department of Mathematics,\\  Khalifa University of Science and Technology,\\ Main Campus, Abu Dhabi,\\ United Arab Emirates}
\author{S. Chanda}
\email{sumanto.chanda@icts.res.in}
\affiliation{
International Centre for Theoretical Sciences\\
No. 151, Shivakote, Hesaraghatta Hobli, Bengaluru, Karnataka 560089, India
}
\author{P. Guha}
\email{partha.guha@ku.ac.ae}
\affiliation{%
Department of Mathematics,\\  Khalifa University of Science and Technology,\\ Main Campus, Abu Dhabi,\\ United Arab Emirates}
\date{\today}
\begin{abstract}
We study optical metrics via null geodesics as a central force system, deduce the related Binet equation and apply the analysis to certain solutions of Einstein's equations with and without spherical symmetry. A general formula for the deflection angle in the weak lensing regime for the Schwarzschild-Tangherlini (ST) metric is derived. In addition, we obtain a new weak lensing formula for the deflection angle on the equatorial plane of a Kerr black hole (BH). We also explore the bending of light by considering the gravitational objects described by the Tomimatsu-Sato (TS) metric.
\end{abstract}

\pacs{04.20.-q, 04.50.Gh, 04.30.Nk, 98.62.Sb}
\keywords{Null geodesic, Optical-mechanical formulation, Binet equation, Schwarzschild-Tangherlini metric, Kerr metric, Tomimatsu-Sato metric, black hole, naked singularity, weak lensing.}

\maketitle



\section{Introduction}
Optical metrics are essentially null geodesics studied by projecting them onto lower dimensional spatial surfaces. If a metric admits a timelike 
Killing vector $K$ orthogonal to a hypersurface, the null geodesic projects down 
to unparameterised geodesics of the optical metric on the space of orbits of $K$. 
Similar constructions were studied for metrics admitting a stationary Killing vector 
\cite{ghww} or a timelike conformal retraction \cite{casey2} where the projected 
null curves provide some notion of geometric structure to a hypersurface. 
Null curve geodesics are formulated according to Fermat's principle, since they 
cannot be minimised. Gravitational lensing is one of the research applications in observational astronomy that can be discussed in terms of optical metrics. Moreover, using null geodesics one can interpret gravitational 
fields as transparent media with a refractive index, and vice versa as already speculated long ago by P. de Fermat and P.L. Maupertuis \cite{maup}. 

The analytical calculations involving null geodesics in spherically symmetric 
space-times using Weierstrass elliptic functions are given in \cite{gwg1,Cie,Hal,Hamo}. Null geodesics can be written in the form of Binet’s equations \cite{casey1}, casting them as central force systems for some potentials. One way to provide solutions to Binet's equation is to solve the null geodesic equation. We do that by deriving a general formula for the weak deflection angle in the case of the ST metric in $n+1$ dimensions \cite{Tangherlini}. In this context, it is worth mentioning that \cite{Eiroa} studied the strong lensing of a braneworld BH modelled in terms of a ST metric with $n=4$ while the weak lensing was investigated by \cite{Majumdar}. The gravitational lensing problem for the general case of $n$ space dimensions was partially addressed by \cite{Tsukamoto} where both strong and weak field limits have been studied. More precisely, the strong lensing was analysed by the method developed by \cite{Bozza} which has been proved to be mathematically ill-defined in \cite{Batic} while the weak lensing has been treated by setting up a recurrence relation for a certain integral which leads to two distinct formulas for the deflection angle depending whether the spatial dimension is even or odd. We showed that the aforementioned integral can be solved in terms of a certain ratio of Gamma functions, thus unifying the treatment of an even/odd number of spatial dimensions. Last but not least, we study the refractive index of the ST metric by using isotropic coordinates. Regarding other applications of the  isotropic coordinates highlighting the effects the refractive index  and the  gravitational lensing may have on the Hawking radiation, we refer to \cite{Sak1,Sak2,Sak3}.

We also dwell into the anisotropic case where the optical anisotropy of curved space is demonstrated by means of a rigorous algebraic analysis in \cite{KR}. In particular, we focus our attention on two space-times: the Kerr metric and the Tomimatsu-Sato manifold with a certain deformation parameter. In the context of a Kerr BH, we derive new formulae for the refractive indices. Moreover, for BHs characterized by a low spin parameter, we construct a novel expression for the deflection angle in the weak lensing regime for light paths restricted to the equatorial plane. Such a result differs from the weak lensing expansions for the deflection angle obtained in \cite{Sereno,Renzini} because it is obtained by exploiting dimensionless parameters different than those used therein. Finally, an exact formula for the deflection angle on the equatorial plane has been derived in terms of complete and incomplete elliptic integrals of the third kind by \cite{Iyer}. However, no weak lensing formula have been offered there. In addition to this, \cite{Ovgun,Khan} followed the optical-geometric approach to estimate the shadow and the deflection angle in the weak lensing regime for a Kerr-Newman-Kasuja BH and a Kerr-Newman BH in quintessential dark energy.

Regarding the TS metric, we shortly recall that it was discovered in 1972 by Tomimatsu and Sato \cite{tomsato, tomsatop} and it describes solutions for stationary axisymmetric system with quadrupole momentum larger than that of a Kerr BH. Such a metric generalises the Zipoy-Voorhes solutions to Einstein's field equation by accounting for the space-time rotation \cite{Zipoy,Voor}. For a discussion of various limits of this metric we refer to \cite{kinkel}. It is worth mentioning that the TS metric, even though it is asymptotically flat, has not been extensively studied as much as other exact solutions to the Einstein field equations. The reason of this lack of studies is probably threefold. From one side, TS space-times exhibit naked singularities and this aspect together with the cosmic censorship may  lead us to think that such space-times are not physically meaningful. However, the cosmic censorship is still a conjecture and up to now, several counterexamples are known \cite{Papa,Hollier,Singh,Gundlach,Harada,Boyda,Israel,Drukker}. On the other side, as already pointed out by \cite{Kodama,Bambi}, it seems that there has been some confusion in the literature regarding the singularity structure of the TS metric. To make this story short, \cite{gary,glass} claimed that the TS solution is a spinning naked singularity without event horizon while thirty years later \cite{Kodama} addressed the conceptual mistakes related to \cite{gary,glass} and showed the existence of a degenerate horizon with two components. Finally, \cite{Bambi} noted that the most pathological behaviour is probably the occurrence of causality violation inside the ring singularity and close to the segment singularity where closed time-like curves can exist. Hence, it seems legitimate to investigate whether the lensing signature of this space-time exhibits some distinct and measurable features when compared with BH systems. A similar investigation but in a different context represented by rotating wormholes was carried out in \cite{Tsu,Chakra,BozzaN,Chakra17,Abdu,Abdu1,Jusufi}. Regarding the study of the light bending in the TS metric, there is a scarcity of results in the literature. For instance, a  partially incorrect weak lensing formula has been derived by \cite{bosewang} while only forty-five years later, \cite{Bambi} studied the shape and position of the shadow for the Tomimatsu-Sato space-time, compared it with that of the Kerr metric and pointed out that future space sub-mm interferometers such as VSOP-3 might be able to distinguish between these gravitational objects. In view of that, we addressed the weak lensing problem on the equatorial plane of the TS metric by an approach differing from that adopted in \cite{bosewang}. Unfortunately, it turns out that the second order correction in the weak lensing formula (31) of \cite{bosewang} is not correct. This motivated us to derive the correct expression for the second order perturbation and to generalise the corresponding formula to higher order corrections. For reasons of space, the strong lensing in the TS metric is not dealt with and it will be considered in another work.

The outline of the paper is as follows. In Section II, we first introduce some preliminary notions on null geodesics, derive Snell's law used in refractive optics, and deduce formulae for the refractive indices. In Section III, we first consider a general family of spherically symmetric space-times, describe the null geodesics 
as a two dimensional central force system, and deduce the related Binet equation. More  precisely, in the context of the ST metric, we derive a general formula representing the deflection angle in the weak field approximation valid at first order in the perturbative parameter and for an arbitrary number of spatial dimensions. The analysis of null geodesics is then applied to other examples of solutions to Einstein's equations such as Kerr space-times and the TS metric. Regarding the Kerr metric, we derive a novel weak lensing approximation for the deflection angle of a null ray moving on the equatorial plane. Concerning the TS metric, we correct and improve the accuracy of a weak lensing formula previously obtained in the literature \cite{bosewang}. In Section IV, we draw our conclusions.

\section{Preliminaries: Null geodesics} \label{sec:sec2}

P.L. Maupertuis speculated in \cite{maup} that light passing through a medium was 
refracted due to gravitational effects. Furthermore, null geodesics are unique since 
the speed of a particle (photon) travelling along a null geodesic remains unchanged 
under local Lorentz group transformations. In special relativity, for flat spaces, this 
leads to Einstein's postulate about the universality of the speed of light in all inertial 
frames, which holds true locally, even in refracting media.
Here, starting with a null geodesic in an isotropic space-time, we will shortly remind the reader how refractive 
phenomena can arise from a gravitational metric (see \cite{br} for further details). Suppose 
we have a space-time metric which in isotropic coordinates is described by the line element
\begin{equation} \label{isotrop} 
ds^2 =g_{\mu\nu}dx^\mu dx^\nu =A (r) dt^2 - B (r)|d\bm{r}|^2,\quad
|d\bm{r}|^2=dr^2+r^2\left(d\vartheta^2+\sin^2{\vartheta}d\varphi^2\right),
\end{equation}
where we adopted geometrized units such that $c=G_N=1$. Taking into account that $ds^2=0$ for null particles and introducing an affine parameterization such that $x^\mu=x^\mu(\lambda)$ yield 
\begin{equation}
\left( \frac{d s}{d \lambda} \right)^2 = A (r) {\dot t}^2 - B (r) |\dot{\bm{r}}|^2=0.
\end{equation}
Here, a dot denotes differentiation with respect to the affine parameter. From the above equation and under the reasonable assumption that $t=t(\lambda)$ is an increasing function of  $\lambda$, it follows that the local refractive index $n(\bm{r})$ with respect to vacuum is
\begin{equation} \label{refindex} 
n (\bm{r}) = \frac{1}{v_{p,n}}= 
\sqrt{\frac{B (r)}{A (r)}},\quad
v_{p,n}=\frac{|\dot{\bm{r}}|}{\dot{t}}=\left|\frac{d\bm{r}}{dt}\right|
\end{equation}
with $v_{p,n}$ denoting the phase velocity of a light ray as measured by a distant observer. Note that in the case of a conformally flat metric $v_{p,n}=1$. Furthermore, if we recast (\ref{refindex}) into the form
\begin{equation}
1=v_{p,n}\sqrt{\frac{B (r)}{A (r)}},
\end{equation}
we can immediately see that at any given position, the speed of light is universal in all inertial frames of reference. It is worth mentioning that in the case of spherically symmetric vacuum solutions to Einstein's equations, $A B = 1$ and the refractive index \eqref{refindex} simplifies as follows 
\begin{equation}
n (\bm{r}) =\frac{1}{|A(r)|}.
\end{equation}
The above scenario changes in the case of an anisotropic space-time because each spatial direction has its own refractive index $n_i (\bf{r})$. More precisely, if we consider the stationary metric with line element    
\begin{equation}\label{anis}
ds^2 = h_{00} (\bm{r}) dt^2 - 2 h_{i0} (\bm{r})d tdx^i - h_{ij} (\bm{r}) dx^i dx^j
\end{equation}
such that for all $i,j=1,2,3$ the metric coefficients $h_{00}$ and $h_{ij}$ are both positive on some common domain $\Omega\subset\mathbb{R}^3$, the refractive index along each spatial direction can be deduced from the following quadratic equation 
\begin{equation}\label{intra}
h_{00} (\bm{r})c^2 - 2h_{i0} (\bm{r})v^i_{p,n} - h_{ii} (\bm{r}) \left( v^i_{p,n} \right)^2 = 0,\quad
v^i_{p,n}=\frac{\dot{x}^i}{\dot{t}}=\frac{dx^i}{dt},
\end{equation}
which has been obtained from (\ref{anis}) with $i=j$ and by setting $ds^2=0$ followed by the introduction of an affine parameterization. If we define the refractive index in the $i$-th spatial direction as
\begin{equation}
n_i ({\bf{r}})=\frac{1}{v^i_{p,n}},
\end{equation}
equation (\ref{intra}) becomes
\begin{equation}
h_{00} (\bm{r}) n_i^2 - 2 h_{i0} (\bm{r}) n_i - h_{ii} (\bm{r}) = 0 
\end{equation}
and the corresponding roots are given by
\begin{equation}\label{rpm}
n_{i,\pm}(\bm{r}) = \frac{h_{i0}(\bm{r}) \pm \sqrt{h_{i0}^2(\bm{r}) + h_{00}(\bm{r}) h_{ii}(\bm{r})}}{h_{00}(\bm{r})}.
\end{equation}
At this point a comment is in order. Since we initially assumed that for all $i,j=1,2,3$ $h_{00},h_{ij}>0$ on $\Omega\subset\mathbb{R}^3$, it is evident that $h_{ii}$ is also positive definite, and the inequality $\sqrt{h_{i0}^2(\bm{r}) + h_{00}(\bm{r}) h_{ii}(\bm{r})} > |h_{i0}(\bm{r})|$ holds true on $\Omega\subset\mathbb{R}^3$. Thus, if $h_{i0}=-|h_{i0}|$, we have a negative refractive index (permissible when considering metamaterials \cite{fnb, cs, csl}), technically signalizing that the group and phase velocities have opposite directions. This situation corresponds to a metamaterial with negative electric and magnetic permeability. Since in the present work we are not interested to model metamaterials that can be used to simulate general-relativity phenomena, we will consider the positive root in \eqref{rpm} which we denote as follows
\begin{equation} \label{nisoindex} 
n_i(\bm{r})=\frac{h_{i0}(\bm{r}) + \sqrt{h_{i0}^2(\bm{r}) + h_{00}(\bm{r}) h_{ii}(\bm{r})}}{h_{00}(\bm{r})}.
\end{equation}
Since the line element of a null geodesic vanishes, we need to vary the spatial part alone. This can be achieved by applying Fermat's principle of light \cite{Synge} stating that  a null ray through the events $1$ and $2$ must obey  the constraints
\begin{equation}
\delta\int_1^2 dt=0,\quad ds=0,
\end{equation}
which can be rewritten in equivalent form as
\begin{equation}
\delta\int_1^2\dot{t}\ d\lambda=0,\quad g_{\mu\nu}\dot{x}^\mu\dot{x}^\nu=0.
\end{equation}
Using \eqref{refindex} and assuming an isotropic space-time, the optical arc integral is
\begin{equation} \label{optlag} 
\ell = \int_1^2 {\dot t} d \lambda= 
\int_1^2  n (\bm{r}) |\dot{\bm{r}}| d \lambda= 
\int_1^2 n (\bm{r}) |d{\bm{r}}| 
\end{equation}
with Lagrangian $\mathcal{L}_{n} = \dot{t} = n (\bm{r}) |\dot{\bm{r}}|$. If we further impose that $\delta\ell=0$, the corresponding Euler-Lagrange equation is
\begin{equation}\label{EL}
\frac{d}{d\lambda}\left(n(\bm{r})\frac{\dot{\bm{r}}}{|\dot{\bm{r}}|}\right) = 
| \dot{\bm{r}}| \bm{\nabla} n (\bm{r}).
\end{equation}
We will regard the arc length $\sigma$ defined through the relation $d\sigma=|\dot{\bm{r}}|d\lambda$ as a natural parameter along the curve. If we parameterize with respect to arc length, the reparameterized velocity can be written as a unit vector $\widehat{\bf{e}}=d\bm{r}/d\sigma=\dot{\bm{r}}/|\dot{\bm{r}}|$ denoting the  direction of the light ray, which in turn allows us to write the Maupertuis action for light-like curves (\ref{optlag}) as
\begin{equation}\label{tramezzo}
\ell = \int_1^2 {\bf{p}}_n\cdot d\bm{r},\quad
{\bf{p}}_n=\frac{\partial\mathcal{L}_n}{\partial\dot{\bm{r}}}=n (\bm{r})\widehat{{\bf{e}}}.
\end{equation}
Taking into account that the time $\Delta\mathcal{T}$ as measured by a distant observer for the light ray to go from $1$ to $2$ is given by 
\begin{equation}
\Delta\mathcal{T}=\int_1^2 \dot{t}d\lambda=\int_1^2 dt=\frac{1}{c}\int_1^2 d\ell,
\end{equation}
and using (\ref{tramezzo}), it is possible to derive as in \cite{cgg} the following Eikonal equation
\begin{equation}\label{eikonal} 
{\bm{\nabla}}\mathcal{T} = {\bm{\nabla}}\ell= {\bf{p}}_n = n({\bm{r}}) {\widehat{\bm{e}}} 
\end{equation}
with $|{\bm{\nabla}}\mathcal{T}|^2=n^2$. Furthermore, we can use the arc length $\sigma$ to rewrite the Euler-Lagrange 
equation (\ref{EL}) as
\begin{equation}\label{imp}
\frac{d}{d\sigma}\left( n (\bm{r}) \widehat{\bm{e}} \right) = 
{\bm{\nabla}} n (\bm{r}) 
\end{equation}
and derive from it the following equation for the light ray 
\begin{equation}\label{triple}
\frac{d {\widehat{\bm{e}}}}{d \sigma} = 
\left( {\widehat{\bm{e}}} \times {\bm{\nabla}} \ln{n} \right) \times {\widehat{\bm{e}}},
\end{equation}
a result which is normally derived from the Eikonal equation (see \cite{br}). These spatial geodesics are better analysed using a Frenet–Serret-like frame as pointed out in \cite{bgj}. We will use this fact to demonstrate that Snell's law is applicable to gravitational fields. To this purpose, we set up a basis $\left( {\widehat{\bm{e}}}_{\parallel}, {\widehat{ \bm{e}}}_{\perp} \right)$ such that ${\widehat{\bm{e}}} = \cos{\theta} \ {\widehat{\bm{e}}}_{\parallel} + \sin{\theta} \ {\widehat{\bm{e}}}_{\perp}$ with $\widehat{\bm{e}}_{\parallel}$ lying along $\bm{\nabla}n$, more precisely $\widehat{\bm{e}}_{\parallel}={\bm{\nabla}} n/|{\bm{\nabla}} n|$, and ${\widehat e}_{\perp}$ is orthogonal to ${\bm{\nabla}} n$. At this point, it is not difficult to verify that
\begin{equation}\label{hilfe}
\frac{d n}{d \sigma} = \frac{d {\bm{r}}}{d \sigma}\cdot {\bm{\nabla}} n = 
{\widehat{\bm{e}}}\cdot{\bm{\nabla}} n = |{\bm{\nabla}} n | \cos{\theta}.
\end{equation}
Since the angle $\theta$ is also parameterized in terms of the arc length $\sigma$, we find in the basis $\left( {\widehat{\bm{e}}}_{\parallel}, {\widehat{ \bm{e}}}_{\perp}\right)$
\begin{equation}\label{rep1}
\frac{d {\widehat{\bm{e}}}}{d \sigma} = - \left( \sin{\theta} \ {\widehat{\bm{e}}}_{\parallel} - 
\cos{\theta} \ {\widehat{\bm{e}}}_{\perp} \right) \frac{d \theta}{d \sigma}.
\end{equation}
Moreover, a lengthy but straightforward manipulation of (\ref{triple}) leads to the following equivalent representation for the rate of change of the unit vector $\widehat{\bm{e}}$, namely
\begin{equation}\label{rep2}
\frac{d {\widehat{\bm{e}}}}{d \sigma}=\sin \theta \left(\sin{\theta} \ {\widehat{\bm{e}}}_{\parallel} - 
\cos{\theta} \ {\widehat{\bm{e}}}_{\perp} \right) \frac{ | {\bm{\nabla}} n|}n.
\end{equation}
If we merge together (\ref{rep1}) and (\ref{rep2}), we end up with the following differential equation 
\begin{equation}\label{anticamera}
\frac{d\theta}{d\sigma}+\frac{|{\bm{\nabla}}n|}{n}\sin{\theta}=0.
\end{equation}
Finally, it is possible to apply (\ref{hilfe}) to (\ref{anticamera}) so that $|{\bm{\nabla}}n|$ can be expressed in terms of the rate of change of the refractive index. In this way, we derive the differential equation
\begin{equation}
\frac{d}{d\sigma}\left(n\sin{\theta}\right)=0,
\end{equation}
which once integrated leads to
\begin{equation} \label{snell} 
n ({\bm{r}}) \sin \theta = const.,
\end{equation}
i.e. the Snell law of refractive optics. This demonstrates that regions with 
gravitational fields are comparable to refractive media. Now, we will study classical 
particle mechanics in optical terms.

\section{Dynamical solutions for null geodesics via Binet's equation}
In Section~\ref{sec:sec2}, we deduced optical principles for null geodesics. Here, we study how Binet's equation derives from null geodesics and in addition, we dynamically compare it to mechanical systems with a central force. In the following, we have also attempted to extend Casey's  work \cite{casey1} on  the ST metric
to other solutions. Without further ado let us recall that a spherically symmetric Lorentzian $(n+1)$-dimensional metric with $S^{n-1}$ symmetry and asymptotically flat can be written as
\begin{equation} \label{spmet} 
\begin{split} 
ds^2 = f(r) dt^2 - \frac{dr^2}{g(r)} - r^2 d \Omega_{n-1}^2 \qquad \text{where }
\end{split} \quad 
\begin{split}
f(r) = 1 + F(r), \quad F(r) = \sum_{i=1}^{\infty} a_i r^{-i} \\ 
g(r) = 1 + G(r), \quad G(r) = \sum_{i=1}^{\infty} b_i r^{-i}
\end{split}.
\end{equation}
In the case $n=3$ and assuming without loss of generality that $\vartheta=\pi/2$, the following directional refractive indices can be easily derived by means of (\ref{nisoindex}) 
\begin{equation} \label{newri}
n_r^2 = \frac1{f(r) g(r)},\quad
n_\varphi^2 = \frac{r^2}{f(r)}.
\end{equation}
The null geodesics are characterized by setting $ds^2 = 0$ in the line element appearing in \eqref{spmet}. Moreover, they can be determined by the extremising process we discussed in Section~\ref{sec:sec2}, i.e. such geodesics can be found by extremising the spatial curve as shown in \eqref{optlag} to take the least time to traverse in accordance with Fermat's principle, with respect to which they can be parameterized. For example, let us impose $ds^2=0$ in \eqref{spmet} for the case $n = 3$ and $\vartheta=\pi/2$. The latter condition can always be imposed because of the spherical symmetry of the manifold. If we further set  $ds_{\mathcal{O}}^2 = dt^2$ in \eqref{spmet}, then the unparameterized geodesics are described in terms of the line element 
\begin{equation} \label{unpara} 
ds_{\mathcal{O}}^2 = \frac{dr^2}{f(r) g(r)} + \frac{r^2}{f(r)} d \varphi^2=
\eta^2 (\rho) \left( d \rho^2 + \rho^2 d \phi^2 \right),
\end{equation}
where in the last step we introduced isotropic coordinates defined in terms of a new radial variable $\rho$ such that
\begin{equation}\label{is_c}
\left(\frac{1}{\rho}\frac{d\rho}{dr}\right)^2=\frac{1}{r^2 g(r)},\quad \eta^2(\rho)=\left(\frac{r}{\rho}\right)^2\frac{1}{f(r)}.
\end{equation}
In the case $\rho$ is an increasing function of $r$, it follows from (\ref{is_c}) that
\begin{equation}
\rho(r)=ke^{\int\frac{dr}{r\sqrt{g(r)}}},\quad \eta(\rho)=\frac{r}{\rho\sqrt{f(r)}}
\end{equation}
with $k$ an arbitrary integration constant. Moreover, the conformal factor $\eta (\rho)$ plays the role of the isotropic refractive index. In case of the Schwarzschild solution where $f(r) = g(r) = 1 - \frac{2 M}r$, the isotropic coordinate in \eqref{unpara} is obtained from the first equation in (\ref{is_c}) which leads to
\begin{equation}\label{prad}
\frac{dr}{\sqrt{r^2 - 2 Mr}}=\pm\frac{d \rho}{\rho}.
\end{equation}
Since we want $\rho\to\infty$ as $r\to\infty$, we need to consider the plus sign in (\ref{prad}). As in \cite{Buch} we  require that $\rho = M/2$ at $r = 2 M$ so that the isotropic coordinate is
\begin{equation}
\rho=\frac{1}{2}\left(r-M+\sqrt{r^2-2Mr}\right),
\end{equation}
which can be inverted to give
\begin{equation}
r=\rho\left(1+\frac{M}{2\rho}\right)^2.
\end{equation}
As we will see below, the condition $\rho(2M)=M/2$ ensures that asymptotically at space-like infinity the refractive index becomes one as we would expect for a vacuum in the absence of any gravitational field. At this point, by means of the second equation in  (\ref{is_c}) we conclude that the conformal factor takes the following form
\begin{equation} \label{iso1} 
\eta (\rho) = \frac{\left(1+\frac{M}{2\rho}\right)^3}{1-\frac{M}{2\rho}}.
\end{equation}
At this point a comment is in order. If we stick to the fact that the refractive index is the ratio between the speed of light in vacuum and its phase velocity in a medium, then slowing down the phase velocity would result in an increase of the refractive index. The latter would diverge in the case of vanishing group velocity signalizing that light is not able to penetrate the medium. This is exactly what happens for the refractive index in (\ref{iso1}) when the isotropic coordinates approaches the event horizon. Hence, to an observer at space-like infinity the event horizon acts like a medium with infinite refractive index. Such a behaviour is not surprising because it is well-known in General Relativity, that even though a light  signal can reach the event horizon in a finite proper time, a distant observer would instead come to the conclusion that the same signal takes an infinite time (as measured by a clock attached to the observer) to cross the event horizon. We shall now discuss solutions to the null geodesic equations that will help to describe BH optics, and apply the formulation developed in Section~\ref{sec:sec2} to other examples among which one that has been already discussed by Casey \cite{casey1} but using a different approach from ours.

\subsection{Central force mechanics and Binet's equation}
We can see that the metric \eqref{spmet} is spherically symmetric, which means that 
we are dealing with mechanical systems governed by central forces. For various force 
laws, the solutions to the equations will describe their respective orbits, allowing us to 
use existing solutions from dynamics to describe null geodesic mechanics. More precisely, the central-force motion can be deduced from the null geodesics of \eqref{spmet}. If we start with the null geodesics of the aforementioned metric upon restriction to motion in the plane $\vartheta = \pi/2$ and we set $n = 3$, the condition $g_{\mu\nu}\dot{x}^\mu\dot{x}^\nu=0$ gives rise to the equation
\begin{equation}\label{conditio}
f(r)\dot{t}^2-\frac{\dot{r}^2}{g(r)}-r^2\dot{\varphi}^2=0,
\end{equation}
where the dot means differentiation with respect to some affine parameter $\lambda$. Since the original metric admits two Killing vectors $\partial_t$ and $\partial_\varphi$, there are two integrals of motion
\begin{equation}\label{conserved}
p_t=\frac{\partial\mathcal{L}}{\partial\dot{t}}=f(r)\dot{t}=E,\quad
p_\varphi=-\frac{\partial\mathcal{L}}{\partial\dot{\varphi}}=r^2\dot{\varphi}=L,
\end{equation}
where $2\mathcal{L}=g_{\mu\nu}\dot{x}^\mu\dot{x}^\nu$ while $E$ and $L$ denote the energy and angular momentum of the massless particle, respectively. Replacing (\ref{conserved}) into (\ref{conditio}) gives rise to the following nonlinear ODE
\begin{equation}\label{kont1}
\frac{f(r)}{g(r)}\left(\frac{dr}{d\lambda}\right)^2+\frac{L^2}{r^2}f(r)=E^2.
\end{equation}
If we introduce a tortoise coordinate $r_*=r_*(r)$ defined through the equation
\begin{equation}
\frac{dr_{*}}{dr}=\sqrt{\frac{f(r)}{g(r)}},
\end{equation}
then, (\ref{kont1}) becomes
\begin{equation}\label{LE}
\left(\frac{dr_{*}}{d\lambda}\right)^2+\frac{L^2}{r^2}f(r)=E^2
\end{equation}
with $r=r(r_*)$. Finally, differentiating the above equation with respect to $\lambda$ yields \begin{equation}\label{eom1}
\frac{d^2 r_{*}}{d\lambda^2}=-\frac{dV_c}{dr_{*}},\quad V_c(r_{*})=\frac{L^2}{2r^2(r_*)}f(r(r_*)).
\end{equation}
Equation (\ref{eom1}) describes a central force system with potential $V_c(r_*)$ for massless particles in spherically symmetric solutions to Einstein field equations. If we consider the tortoise coordinate $r_*$ as a function of the azimuthal angle $\varphi$, then by means of the second relation in (\ref{conserved}) we find that
\begin{equation}
\frac{dr_*}{d\lambda}=\frac{L}{r^2}\sqrt{\frac{f(r)}{g(r)}}\frac{dr}{d\varphi},
\end{equation}
which in turn allows us to rewrite (\ref{LE}) as
\begin{equation}
\frac{L^2}{r^4}\frac{f(r)}{g(r)}\left(\frac{dr}{d\varphi}\right)^2+\frac{L^2}{r^2}f(r)=E^2.
\end{equation}
Finally, setting $u=1/r$ and introducing the impact parameter $1/b^2=E^2/L^2$ leads to the following equation 
\begin{equation}\label{interludio}
\frac{\widetilde{f}(u)}{\widetilde{g}(u)}\left(\frac{du}{d\varphi}\right)^2+u^2\widetilde{f}(u)=\frac{1}{b^2},
\end{equation}
which can be cast into the form
\begin{equation}\label{dyn_m}
\left(\frac{du}{d\varphi}\right)^2+u^2=-u^2\widetilde{G}(u)+\frac{1}{b^2}\frac{\widetilde{g}(u)}{\widetilde{f}(u)}
\end{equation}
whith $\widetilde{G}=G(1/u)$, $\widetilde{f}=f(1/u)$ and $\widetilde{g}=g(1/u)$. Equation (\ref{dyn_m}) describes the geometry of the null geodesics in the invariant plane. Taking into account that as a result of the transformation $u=1/r$ on the metric (\ref{spmet}) the refractive indices of the transformed metric are 
\begin{equation}
\widetilde{n}_r^2=\frac{1}{u^4\widetilde{f}(u)\widetilde{g}(u)},\quad
\widetilde{n}_\varphi^2=\frac{1}{u^2\widetilde{f}(u)},
\end{equation}
equation (\ref{interludio}) can be rewritten in terms of the refractive indices (\ref{newri}) as
\begin{equation}\label{mde} 
\left(\frac{d u}{d \varphi} \right)^2= \frac{\widetilde{n}^2_\varphi}{\widetilde{n}^2_r}\left(\frac{\widetilde{n}^2_\varphi}{b^2}-1\right),
\end{equation}
where a tilde indicates that the radial variable $r$ must be understood as a function of $u$.
A short comment is in order. First of all, we observe that motion reality condition is $\widetilde{n}_\varphi>b$. Moreover, it is gratifying to see that equation (\ref{mde}) correctly reproduces the corresponding one for the Schwarzschild case (see equation (214) p. 123 in \cite{Chandra}). To get the Binet equation, it is more convenient to differentiate (\ref{interludio}) with respect to $\varphi$. Hence, we obtain
\begin{equation}\label{nuova}
\frac{d^2 u}{d\varphi^2}+u\frac{\widetilde{g}(u)}{\widetilde{f}(u)}+\frac{1}{2}\frac{d\ln{(\widetilde{f}(u)/\widetilde{g}(u))}}{du}\left(\frac{du}{d\varphi}\right)^2=\frac{\widetilde{g}(u)}{2\widetilde{f}(u)}\frac{d}{du}\left[u^2(1-\widetilde{f}(u))\right].
\end{equation}
As a consistency check, we observe that for the subclass of metrics (\ref{spmet}) with $f=g$, equation (\ref{nuova}) reduces to
\begin{equation}\label{vv}
\frac{d^2 u}{d\varphi^2}=-\frac{1}{2}\frac{d}{du}\left(u^2\widetilde{f}(u)\right).
\end{equation}
If we further choose $\widetilde{f}(u)=1-2Mu$, then (\ref{vv}) correctly reproduces the Binet equation (32) with $\epsilon=0$ in \cite{Hernandez} for the case of the Schwarzschild metric. Finally, the expression for the Binet equation involving the refractive indices (\ref{newri}) can be obtained from (\ref{interludio}) or equivalently from (\ref{nuova}). More precisely, it is given by
\begin{equation}\label{binet}
\frac{d^2 u}{d\varphi^2}+\frac{d}{du}\left(\frac{\widetilde{n}_r}{u^2\widetilde{n}^2_\varphi}\right)\left(\frac{du}{d\varphi}\right)^2=-\frac{1}{2}\frac{d}{du}\left(\frac{1}{\widetilde{n}^2_\varphi}\right).
\end{equation}
If we choose to work with the isotropic counterpart of the metric (\ref{spmet}), i.e.
\begin{equation}
ds^2 = A(\rho) dt^2 - B(\rho)\left[d\rho^2 + \rho^2(d\vartheta^2+\sin^2{\vartheta}d\varphi^2)\right],
\end{equation}
the refractive indices are found to be
\begin{equation}
n_\rho^2 =\frac{B(\rho)}{A(\rho)},\quad n_\varphi^2 =\rho^2 n_\rho^2.
\end{equation}
Since the procedure to derive the Binet equation is similar to that outlined above, we will limit us here to provide only the main results. Taking into account that in the present case
\begin{equation}
\dot{t}=\frac{E}{A(\rho)},\quad\dot{\varphi}=\frac{L}{\rho^2 B(\rho)},
\end{equation}
and setting $ds^2=0$ yields
\begin{equation}\label{letzte}
A(\rho)B(\rho)\left(\frac{d\rho}{d\lambda}\right)^2+\frac{L^2 A(\rho)}{\rho^2 B(\rho)}=E^2.
\end{equation}
and introducing the tortoise coordinate
\begin{equation}
\frac{d\rho_*}{d\rho}=\frac{1}{\sqrt{A(\rho)B(\rho)}},
\end{equation}
we find that 
\begin{equation}
\frac{d^2\rho_*}{d\lambda^2}=-\frac{d\widetilde{V}_c}{d\rho_*},\quad
\widetilde{V}_c=-\frac{L^2}{2\rho_*^2}\frac{A(\rho)}{B(\rho)},
\end{equation}
where $\widetilde{V}_c$ denotes the associated central potential. Assuming $\rho=\rho(\varphi)$ and by means of the coordinate transformation $\rho=1/u$ applied to (\ref{letzte}) leads to the equivalent equation
\begin{equation} \label{binet2} 
\left( \frac{d u}{d \phi} \right)^2 + u^2 = \frac{1}{b^2}\frac{\widetilde{B}(u)}{\widetilde{A}(u)},
\end{equation}
where $b$ denotes as usual the impact parameter and $\widetilde{A}=A(1/u)$, $\widetilde{B}=B(1/u)$. The above equation can be compactly written in terms of the refractive index $\widetilde{n}_\rho$ as follows
\begin{equation}
\left( \frac{d u}{d \phi} \right)^2 + u^2 = \left(\frac{\widetilde{n}_\rho}{b}\right)^2.
\end{equation}
Finally, in order to obtain the corresponding Binet equation, we first multiply (\ref{binet2}) by $\widetilde{A}/\widetilde{B}$ and differentiate with respect to $\varphi$. This leads to the ODE
\begin{equation}
\frac{d^2 u}{d\varphi^2}+u-\frac{d\ln{\widetilde{n}_\rho}}{du}\left(\frac{du}{d\varphi}\right)^2=\frac{d\ln{\widetilde{n}_\rho}}{du}.
\end{equation}
Thus, solutions to Binet's equations available in dynamics should help  describe the null geodesic trajectories for various force laws. 

\subsection{Applications}

\cite{casey1} studied optical metrics via null geodesics of the ST solution. Here, starting with the more general metric (\ref{spmet}) which contains the ST metric as a special case, we shall reproduce the Binet equation deduced by \cite{casey1} (see equation (2) therein) and focus on the optical and 
dynamical properties of the ST manifold. In the next section, we shall extend the results in \cite{casey1} to other solutions. First of all, we observe that the ST metric follows from (\ref{spmet}) by setting $f(r)=g(r)$, i.e. $a_j=b_j$ for all $j\in\mathbb{N}$  and fixing the aforementioned coefficients so that $f(r)=1-(2M_n/r^{n-2})$. The refractive indices for the ST metric are
\begin{equation}
n_r=\frac{1}{1-\frac{2M_n}{r^{n-2}}},\quad n^2_\varphi=r^2 n_r.
\end{equation}
By means of the transformation $u=1/r$ we have $\widetilde{f}(u)=1-2M_n u^{n-2}$ and from (\ref{dyn_m}) we obtain in agreement with \cite{casey1} the following ODE 
\begin{equation}\label{mde3} 
\left( \frac{d u}{d \varphi} \right)^2 + u^2 = 2 M_n u^n + \frac1{b^2}.
\end{equation}
If we differentiate (\ref{mde3}) with respect to $\varphi$, we end up with the Binet equation
\begin{equation}
\frac{d^2 u}{d \varphi^2} + u = n M_n u^{n-1} .
\end{equation}
Finally, we observe that being $f=g$ in the ST metric, there is no need to introduce a tortoise coordinate and therefore, the central potential and force can be obtained directly from (\ref{eom1}). More precisely, we find
\begin{equation}
V_c(r)=\frac{L^2}{2}\left(\frac{1}{r}^2-\frac{2M_n}{r^n}\right),\quad
F_c(r)=-\frac{dV_c}{dr}=\frac{L^2}{r^3}-\frac{nM_n L^2}{r^{n+1}}.
\end{equation}

\subsection{Solutions of Binet's equations}
Using equation (\ref{mde3}), it should be possible to deduce the deflection angle $\Delta\varphi$ 
by quadrature. Taking the square root of (\ref{mde3}) where without loss of generality we take the plus sign which describes light rays approaching the gravitational source on the equatorial plane along trajectories followed along an anticlockwise direction, we end up with the first order nonlinear ODE
\begin{equation}\label{oode}
\frac{d\varphi}{du}=\frac{1}{\sqrt{\frac{1}{b^2}-u^2\widetilde{f}(u)}}.
\end{equation}
On the other hand, $du/d\varphi$ vanishes at the distance of closest approach $u_0$. This observation allows to express the impact parameter in terms of the metric components according to the relation
\begin{equation}
\frac{1}{b^2}=u_0^2\widetilde{f}(u_0).
\end{equation}
Hence, we can rewrite (\ref{oode}) as
\begin{equation}
\frac{d\varphi}{du}=\frac{1}{\sqrt{u_0^2\widetilde{f}(u_0)-u^2\widetilde{f}(u)}}.
\end{equation}
Before integrating the above equation, we observe that the overall variation $\varphi$ undergoes as $u$ ranges from $0$ (i.e. $r=\infty$) to $u_0$ and from there back to $0$ is $2\left|\varphi(u_0)-\varphi(0)\right|$ which would be exactly $\pi$ if no gravitational object were not present. Hence, as in \cite{Weinberg} we conclude that the angle $\Delta\varphi$ capturing the deviation of light trajectory from a straight line is given by
\begin{equation}
\Delta\varphi=2\left|\varphi(u_0)-\varphi(0)\right|-\pi=2\int_0^{u_0}\frac{du}{\sqrt{u_0^2\widetilde{f}(u_0)-u^2\widetilde{f}(u)}}-\pi.
\end{equation}
Finally, by means of the variable substitution $w=u/u_0$ and introducing the small parameter $h_n=M_n u_0^{n-2}$ (which corresponds to the usual small parameter $M/r_0$ for $n=3$), the computation of the bending of light boils down to the evaluation of the following integral
\begin{equation}\label{anform}
\Delta\varphi_n=2\int_0^1\frac{dw}{\sqrt{1-2h_n-w^2+2h_n w^n}}-\pi
\end{equation}
with $n\in\mathbb{N}$ and $n\geq 3$. Note that we underlined the dependence of the deflection angle on the spatial dimension $n$ by adding the subscript $n$ to $\Delta\varphi$. At this point, it is gratifying to see that (\ref{anform}) correctly reproduces for $n=3$ equation (20) in \cite{Keeton}, i.e. the formula for the deflection angle of light rays propagating in the Schwarzschild metric. In the case $n=3$, the integral appearing in (\ref{anform}) can be computed by means of the formula $234.00$ in \cite{Byrd}. The deflection angle is then expressed in terms of an elliptic integral of the first kind as follows
\begin{equation}
\Delta\varphi_3=\frac{4F(\alpha,k)}{\sqrt[4]{F(h_3)}}-\pi,\quad F(h_3)=1+4h_3-12h_3^2
\end{equation}
with
\begin{equation}
\alpha=\arcsin{\sqrt{\frac{8h_3\sqrt{F(h_3)}}{(6h_3-1+\sqrt{F(h_3)})(1-2h_3+\sqrt{F(h_3)})}}},\quad
k^2=\frac{6h_3-1+\sqrt{F(h_3)}}{2\sqrt{F(h_3)}}.
\end{equation}
At this point, the expansion of $\Delta\varphi_3$ in the small parameter $h_3$ correctly reproduces the weak lensing result for the Schwarzschild metric, namely
\begin{equation}
\Delta\varphi_3=4h_3+\mathcal{O}(h_3^2).
\end{equation}
The case $n=4$ can also be solved analytically. By means of the transformation $w^2=z$ the quartic polynomial appearing under the square root in (\ref{anform}) can be transformed into a cubic which in turn allows to express the final result in terms of a complete elliptic integral of the first kind. More precisely, we have
\begin{equation}
\Delta\varphi_4=\int_0^1\frac{dz}{\sqrt{z(2h_4 z^2-z-2h_4+1)}}-\pi.
\end{equation}
Since the roots of the quadratic are located at $1$ and $(1-2h_4)/2h_4\gg 1$, we can immediately apply 3.131.4 in \cite{Grad} to get 
\begin{equation}
\Delta\varphi_4=\frac{2}{\sqrt{1-2h_4}}K\left(\sqrt{\frac{2h_4}{1-2h_4}}\right)-\pi
\end{equation}
and the corresponding angle in the weak lensing regime is
\begin{equation}
\Delta\varphi_4=\frac{3}{2}\pi h_4+\mathcal{O}(h_4^2).
\end{equation}
For $n\geq 5$ the integral in (\ref{anform}) is hyperelliptic making its exact computation troublesome. In the weak lensing case, we can expand the integrand with respect to the small parameter $h_n$ to get
\begin{equation}\label{wl}
\Delta\varphi_n=-\pi+2\int_0^1\frac{dw}{\sqrt{1-w^2}}+2h_n\int_0^1\frac{1-w^n}{(1-w^2)^{3/2}}dw+\mathcal{O}(h_n^2).
\end{equation}
The first integral contributes by $\pi/2$ while the second integral can be computed with Mathematica to be
\begin{equation}
\int_0^1\frac{dw}{\sqrt{1-w^2}}+2h_n\int_0^1\frac{1-w^n}{(1-w^2)^{3/2}}dw=\sqrt{\pi}\frac{\Gamma\left(\frac{1+n}{2}\right)}{\Gamma\left(\frac{n}{2}\right)}.
\end{equation}
Hence, the deflection angle for weak lensing is
\begin{equation}
\Delta\varphi_n=2\sqrt{\pi}\frac{\Gamma\left(\frac{1+n}{2}\right)}{\Gamma\left(\frac{n}{2}\right)}h_n+{O}(h_n^2).
\end{equation}
It is straightforward to verify that the above formula correctly reproduces the previously obtained  weak lensing angles for $n=3$ and $4$. We shall now examine the Binet's equations formulated for other metrics.

\subsubsection{Examples of Helmholtz and Helmholtz-Duffing oscillators in General Relativity}
Most solutions to the Einstein equations for spherically symmetric space-times will have $f(r) = g(r)$. Here, we have considered two such examples: the Schwarzschild-de Sitter or Kottler metric and the Reissner-Nordstr\"{o}m-de Sitter metric, that is
\begin{eqnarray}
f_H (r) &=& 1 - \frac{2M}r-\frac{\Lambda}{3}r^2, \\
f_{HD} (r) &=& 1-\frac{2M}r- \frac Q{r^2}-\frac{\Lambda}{3}r^2,
\end{eqnarray}
where $Q$ and $\Lambda$ are the BH charge and the cosmological constant, respectively. According to \eqref{eom1}, the central forces and potentials for these solutions are computed to be
\begin{eqnarray}
V_H(r)&=&\frac{L^2}{2}\left(\frac{1}{r^2}-\frac{2M}{r^3}\right)-\frac{L^2\Lambda}{6},\quad
~~~~~~~~~~F_H(r)=L^2\left(\frac{1}{r^3}-\frac{3M}{r^4}\right),\\
V_{HD}(r)&=&\frac{L^2}{2}\left(\frac{1}{r^2}-\frac{2M}{r^3}+\frac{Q^2}{r^4}\right)-\frac{L^2\Lambda}{6},\quad
F_{HD}(r)=L^2\left(\frac{1}{r^3}-\frac{3M}{r^4}+\frac{2Q^2}{r^5}\right).
\end{eqnarray}
The refractive indices for these space-times can be readily obtained from (\ref{nisoindex})
\begin{eqnarray}
\widetilde{n}_{rH} (u) &=& \left( 1 - 2Mu-\frac{\Lambda}{3u^2} \right)^{-1},\quad
~~~~~~~~~~~\widetilde{n}_{\varphi H} (u)=\left(u^2-2Mu^3-\frac{\Lambda}{3}\right)^{-1/2},\\
\widetilde{n}_{rHD} (u) &=& \left( 1 - 2Mu+Qu^2-\frac{\Lambda}{3u^2} \right)^{-1},\quad
\widetilde{n}_{\varphi HD} (u)=\left(u^2-2Mu^3+Q^2u^4-\frac{\Lambda}{3}\right)^{-1/2}.
\end{eqnarray}
With the help of (\ref{binet}) the corresponding Binet equations are given as
\begin{eqnarray}
\label{hmhtz} 
\frac{d^2 u}{d \phi^2} + u &=& 3Mu^2, \\
\label{helmduff} 
\frac{d^2 u}{d \phi^2} + u &=& 3Mu^2 - 2Q^2u^3.
\end{eqnarray}
They are equivalent to the equations for the Helmholtz oscillator \cite{almsan} and the 
Helmholtz-Duffing oscillator \cite{asyyky} respectively, both of which are nonlinear equations that have received a lot of attention recently for the wide range of applications in engineering. The 
solution of the Helmholtz equation \eqref{hmhtz} is given in terms of the Jacobi elliptic function $sn$.  The exact solution of the Helmholtz-Duffing oscillator equation \eqref{helmduff} can also be expressed in 
terms of Jacobi elliptic function \cite{cvet,cvet1}. In this context, it should be noted that Gibbons and Vyska \cite{gwg1} used 
Weierstrass elliptic functions to give a full description and classification of null geodesics in the Schwarzschild space-time.

\subsubsection{Kerr metric}
The black-hole space-time known as the rotating (Kerr) BH is a stationary metric whose line element in geometrized units $c =G_N= 1$ can be expressed in Boyer-Lindquist coordinates $(t,r,\vartheta,\varphi)$ as \cite{Rez}
\begin{equation}
d s^2 =\left( 1-\frac{2Mr}{\Sigma}\right)dt^2+\frac{4Mar\sin^2{\vartheta}}{\Sigma} dtd\varphi-\frac{\Sigma}{\Delta}dr^2
-\Sigma d\vartheta^2-\left(r^2+a^2+\frac{2Ma^2 r}{\Sigma}\sin^2{\vartheta}\right)\sin^2{\vartheta}d\varphi^2
\end{equation}
with
\begin{equation}
\Delta (r)=r^2- 2Mr + a^2,\quad \Sigma=r^2+a^2\cos^2{\vartheta}
\end{equation}
and $a=J/M$ denoting the angular momentum per unit mass of the BH. Using the formulation for non-isotropic spaces introduced in Sec.~\ref{sec:sec2} and in particular, equation  \eqref{nisoindex}, we have the following refractive indices
\begin{eqnarray}
n_r &=& \frac{\Sigma}{\sqrt{\Delta\left(\Sigma - 2Mr\right)}}, \quad n_\vartheta = \frac{\Sigma}{\sqrt{\Sigma - 2Mr}}, \\
n_\varphi &=&\frac{\sin{\vartheta}}{\Sigma-2Mr}\left(4Mar\sin{\vartheta}+\sqrt{\Delta\Sigma^2+12M^2a^2 r^2\sin^2{\vartheta}}\right).
\end{eqnarray}
A this point, a remark about the reality of the roots appearing in the above expressions is necessary. First of all, the inequality $\Sigma-2Mr\geq r(r-r_{h,S})$ holds for all $\vartheta\in[0,\pi]$ where $r_{h,S}$ denotes the event horizon of a Schwarzschild BH. Moreover, the event horizon of a Kerr BH $r_h=M+\sqrt{M^2+a^2}$ satisfies the inequality $r_h>r_{h,S}$. This allows us to conclude that $\Sigma-2Mr>0$ for all $r>r_{h}$ and all $\vartheta\in[0,\pi]$. Let us consider a null ray on the equatorial plane $\vartheta=\pi/2$ where the metric simplifies to
\begin{equation}
ds^2_e = \left( 1 - \frac{2M}r \right)dt^2 + \frac{4M a}r d t d\varphi - \frac{r^2}{\Delta} dr^2 - \left( r^2 + a^2 + \frac{ 2M a^2}r \right) d\varphi^2
\end{equation}
and the associated Lagrangian reads
\begin{equation}
\mathcal{L} =\frac{1}{2}\left(\frac{d\sigma_e}{d\lambda}\right)^2 = \frac{1}{2}\left(1-\frac{2M}r\right)\dot t^2+\frac{2M a}r\dot t \dot\varphi-\frac{r^2}{2\Delta} \dot r^2 -\frac{1}{2}\left(r^2 +a^2+\frac{2Ma^2}r\right)\dot\varphi^2
\end{equation}
with $\lambda$ some affine parameter. The stationarity and axisymmetry of the metric lead to the following two conserved quantities \cite{Chandra}
\begin{equation} \label{kerrcons}
\left(1-\frac{2M}r\right)\dot{t}+\frac{2Ma}{r}\dot{\varphi}=E,\quad
-\frac{2Ma}{r}\dot{t}+\left(r^2 +a^2+\frac{2Ma^2}r\right)\dot\varphi=L_z,
\end{equation}
where $E$ is the energy of the massless particle and $L_z$ the projection of its angular momentum along the BH spin axis. In the following, we are interested in the case $E>0$ so the distinction between direct ($L_z>0$) and retrograde orbits ($L_z<0$) is controlled by the sign of $L_z$. From \eqref{kerrcons}, we can deduce that
\begin{equation}\label{tf}
\dot{t}=\frac{1}{\Delta}\left[\left(r^2 +a^2+\frac{2Ma^2}r\right)E-\frac{2Ma}{r}L_z\right],\quad
\dot{\varphi}=\frac{1}{\Delta}\left[\left(1-\frac{2M}{r}\right)L_z+\frac{2Ma}{r}E\right].
\end{equation}
Finally, substituting the above equations into the null-condition $g_{\mu\nu}\dot{x}^\mu\dot{x}^\nu=0$ with $\vartheta=\pi/2$ yields
\begin{equation}\label{raggio}
\dot{r}^2=E^2+\frac{2M}{r^3}\left(Ea-L_z\right)^2+\frac{E^2a^2-L_z^2}{r^2}.
\end{equation}
The corresponding central force and potential on the equatorial plane are
\begin{equation}\label{fp}
F_{Kerr}(r)=\frac{L^2_z-E^2 a^2}{r^3}-\frac{3M}{r^4}(Ea-L_z)^2,\quad
V_{Kerr}(r)=\frac{L^2_z-E^2a^2}{2r^2}-\frac{M}{r^3}(Ea-L_z)^2.
\end{equation}
From the above equations, we immediately see that orbits on the equatorial plane with $L_z=Ea$ are special in the sense that the solutions $t$ and $\varphi$ to (\ref{tf}) blows up as as the light ray approaches the event horizon \cite{Chandra}. This phenomenon already appears in the context of Schwarzschild and Reissner-Nordstro\"{o}m BHs. Let us introduce the impact parameter $b=L_z/E$. From the potential in (\ref{fp}) it is trivial to infer the existence of an unstable circular orbit for a certain critical value $b_c$ of the impact parameter at
\begin{equation}\label{radius}
r_c=3M\frac{b_c-a}{b_c+a}.
\end{equation}
Note that (\ref{radius}) reproduces the radius of the photon sphere in the limit $a\to 0$. In what follows we are interested in the case of trajectories with $b>b_c$. By means of the second equation in (\ref{tf}) coupled with (\ref{raggio}) we find that
\begin{equation}\label{kerrtraj} 
\left( \frac{d u}{d\varphi} \right)^2 =\left[\frac{1}{b^2}+2M\left(\frac{a}{b}-1\right)^2 u^3+\left(\frac{a^2}{b^2}-1\right)u^2\right]\left(\frac{1-2Mu+a^2 u^2}{1-2Mu+\frac{2Ma}{b}u}\right)^2
\end{equation}
from which the following Binet equation is derived
\begin{equation} \label{binetkerr} 
\frac{d^2 u}{d \varphi^2} = \frac{d W}{du},\quad 
W (u) = \frac{1}{2}\left[\frac{1}{b^2}+2M\left(\frac{a}{b}-1\right)^2 u^3+\left(\frac{a^2}{b^2}-1\right)u^2\right]\left(\frac{1-2Mu+a^2 u^2}{1-2Mu+\frac{2Ma}{b}u}\right)^2.
\end{equation}
We assume that both the observer and the light source are on the equatorial plane. Moreover, we suppose that the light ray follows a trajectory along an anticlockwise direction. Taking into account that at the distance of closest approach $r_0=1/u_0$ 
\begin{equation}
\left.\frac{du}{d\varphi}\right|_{u=u_0}=0,
\end{equation}
the following quadratic equation emerges for the impact parameter
\begin{equation}
1+2M\left(a-b\right)^2 u_0^3+\left(a^2-b^2\right)u_0^2=0
\end{equation}
whose roots are 
\begin{equation}
b_{\pm}=-\frac{2Mau_0^2\pm\sqrt{1-2Mu_0+a^2 u_0^2}}{u_0(1-2Mu_0)}.
\end{equation}
As in \cite{Iyer}, we observe that the sign of the impact parameter encodes the information whether $L_z$  is in the same or opposite direction to the spin $a$. Then, from (\ref{kerrtraj}) we find that the deflection angle is given by the following formula
\begin{equation}\label{angolokerr}
\Delta\varphi_{\pm}=-\pi+2\int_0^{u_0}\frac{du}{\sqrt{\frac{1}{b_\pm^2}+2M\left(\frac{a}{b_\pm}-1\right)^2 u^3+\left(\frac{a^2}{b_\pm^2}-1\right)u^2}}\frac{1-2Mu+\frac{2Ma}{b_\pm}u}{1-2Mu+a^2 u^2}.
\end{equation}
Let us make the change of variable $w=u/u_0$ and introduce the two adimensional parameters $h=Mu_0$, $\xi=au_0$. At this point a remark is in order. For the weak lensing regime $r_0\gg 1$ and hence, $h$ is a small parameter. On the other hand, we also have $u_0\ll 1$ and if we additionally assume that the BH is slowly rotating, i.e. $a\ll 1$, we can conclude that $\xi$ can also be regarded as a small parameter. Hence, we rewrite (\ref{angolokerr}) as
\begin{equation}\label{let}
\Delta\varphi_{\pm}=-\pi+2\int_0^1 F_{\pm}(h,\xi,w)dw
\end{equation}
with
\begin{equation}
F_{\pm}(h,\xi,w)=\frac{1}{\sqrt{\frac{1}{\widehat{b}_\pm^2}+2h\left(\frac{\xi}{\widehat{b}_\pm}+1\right)^2 w^3+\left(\frac{\xi^2}{\widehat{b}_\pm^2}-1\right)w^2}}\frac{1-2hw-\frac{2h\xi}{\widehat{b}_\pm}w}{1-2hw+\xi^2 w^2},\quad
b_\pm=-\frac{\widehat{b}_\pm}{u_0},\quad
\widehat{b}_\pm=\frac{2h\xi\pm\sqrt{1-2h+\xi^2}}{1-2h}.
\end{equation}
Applying Taylor's theorem for multivariate functions to $F_{\pm}$ and integrating over the range of the variable $w$ in (\ref{let}) leads to the following formula for the deflection angle
\begin{equation}\label{Kerrab}
\Delta\varphi_{\pm}=4h+\left(\frac{15}{4}\pi-4\right)h^2\pm 2h\xi+\left(\frac{122}{3}-\frac{15}{2}\pi\right)h^3+\frac{2}{3}h\xi^2\pm\frac{2}{3}(10\pi-16)h^2\xi+\cdots,
\end{equation}
where the BH rotation enters as a second order effect. It is worth observing that the above formula was not derived in \cite{Iyer,Sereno} and it differs from the weak lensing approximation for the deflection angle provided by \cite{Renzini} where the expansion is performed with respect to the parameters $M/b$ and $a/M$.

\subsection{Null geodesics of the TS metric with $\delta=2$}
\begin{figure}
    \centering
    \includegraphics[width=0.3\textwidth]{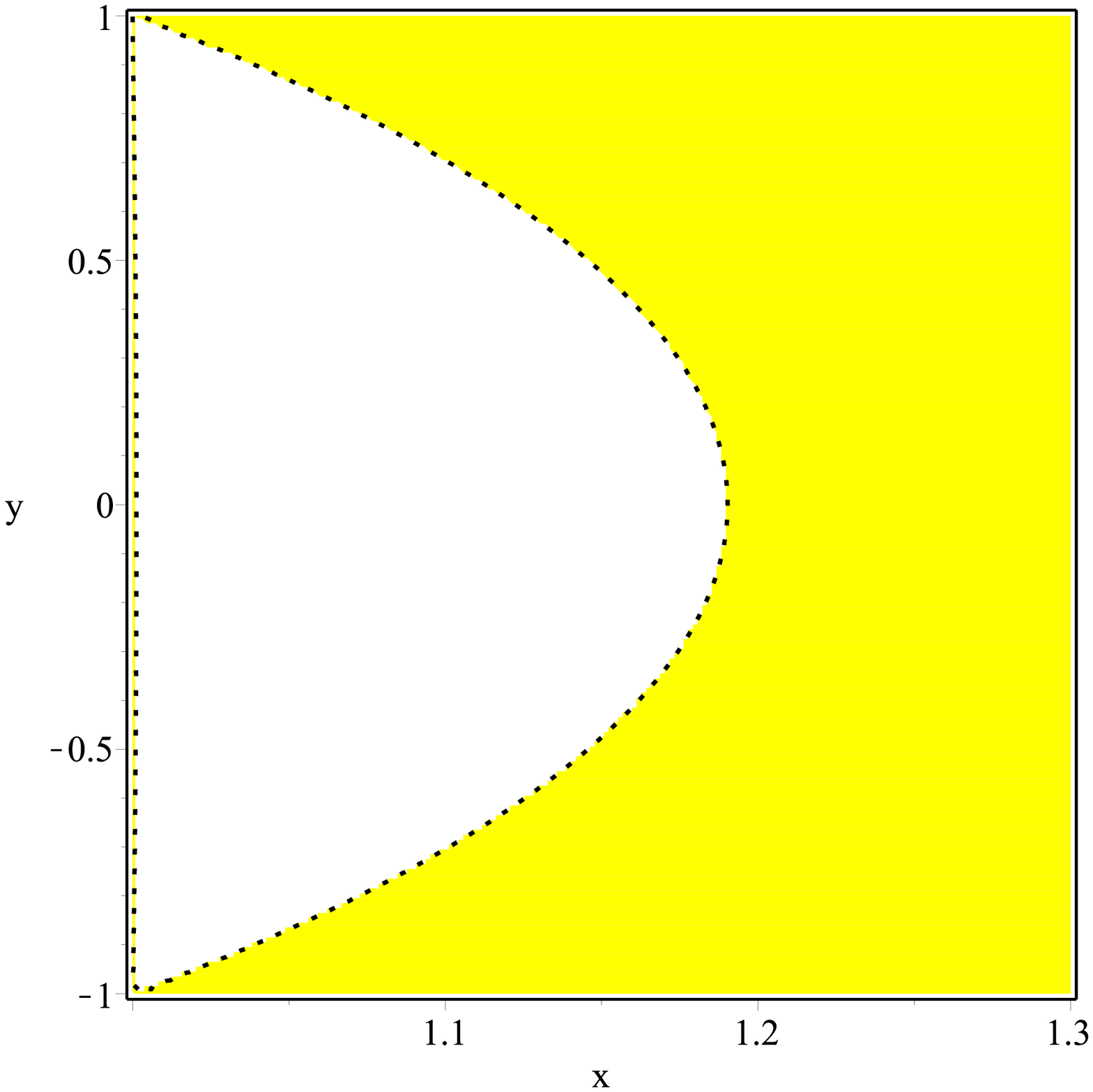}
    \includegraphics[width=0.3\textwidth]{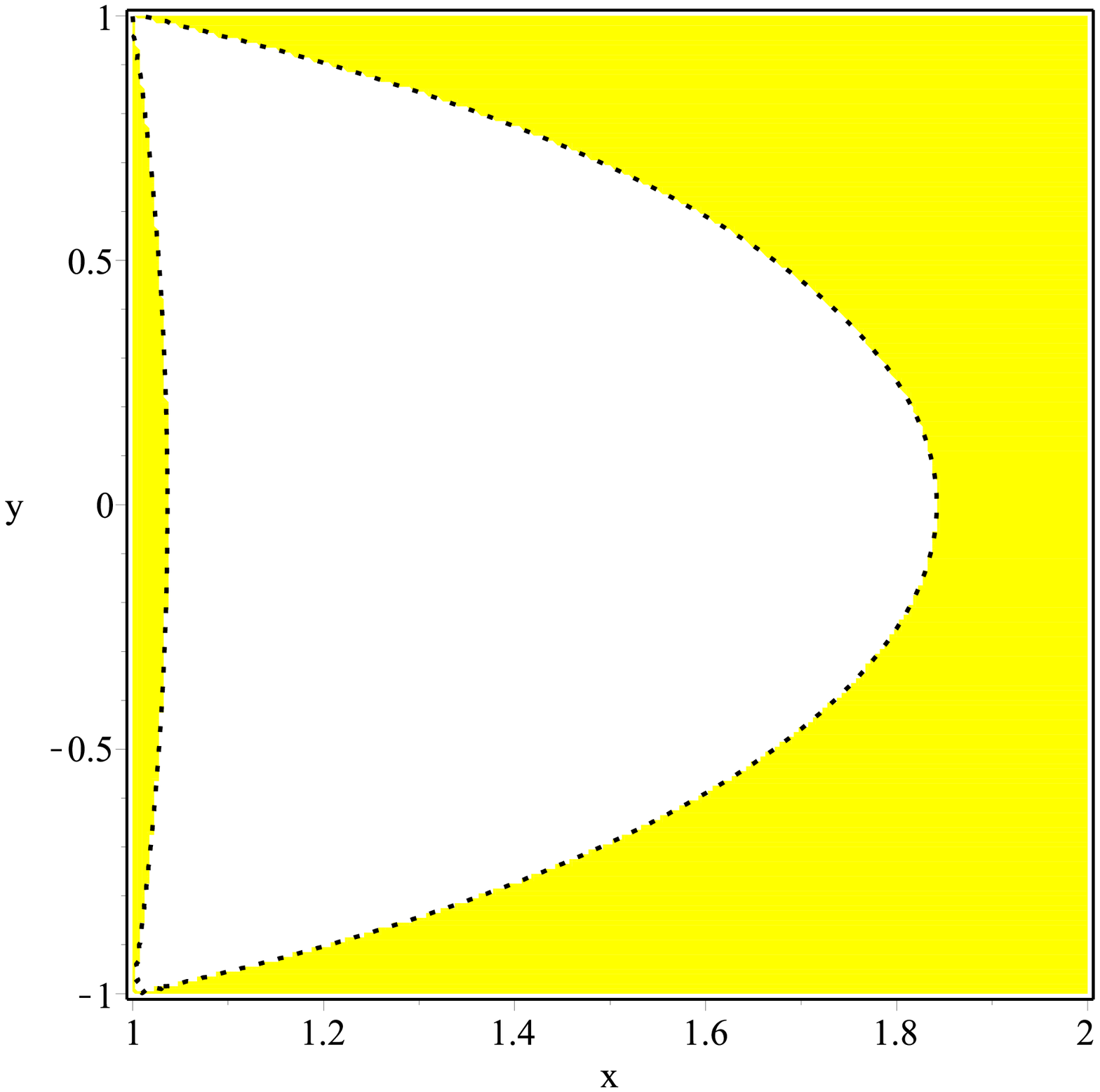}
    \includegraphics[width=0.3\textwidth]{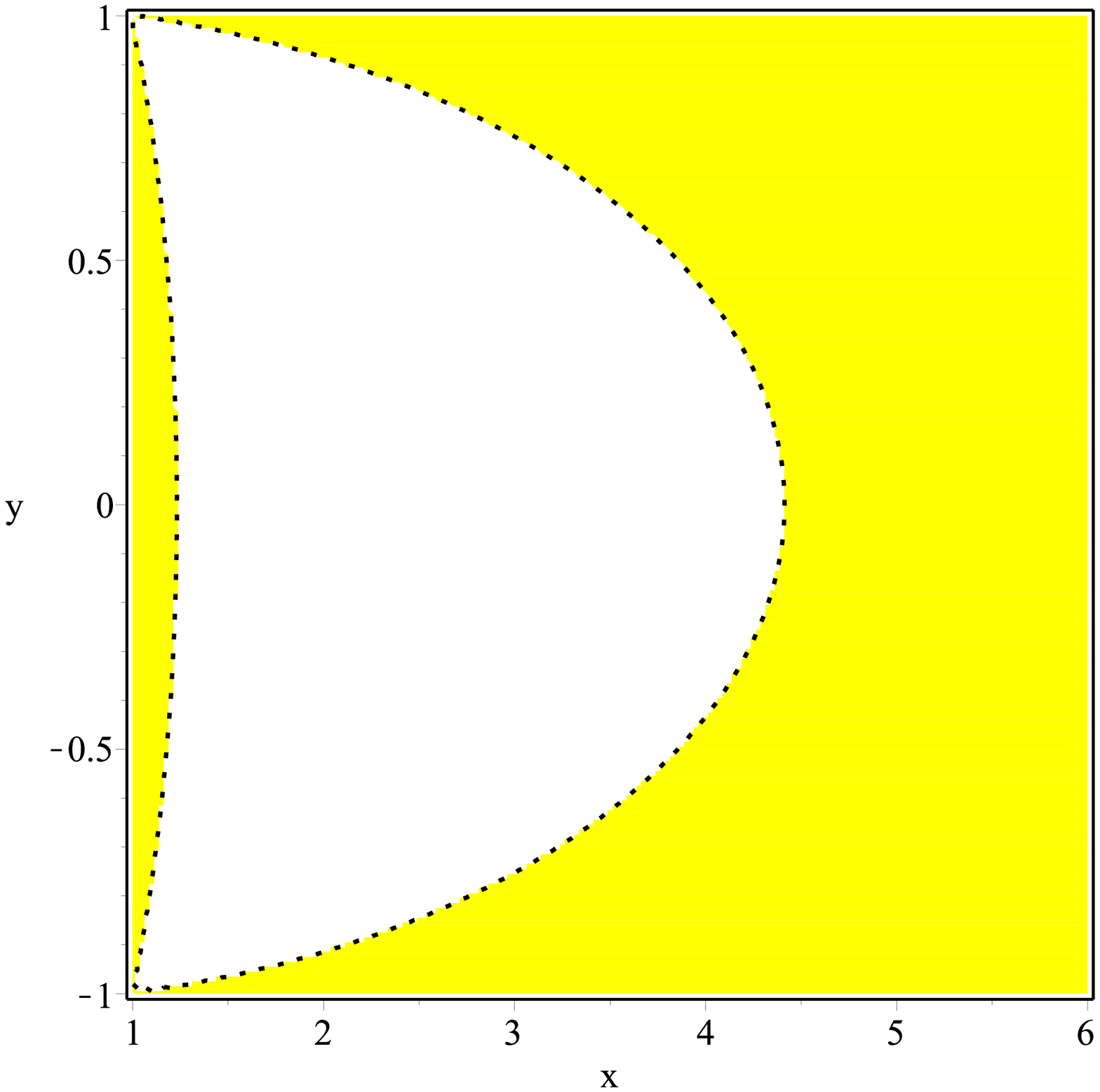}
    \caption{The yellow regions represents those portions of the $(x,y)$-plane where the inequality $A>0$ with $A$ as in (\ref{Alungo}) is satisfied. From left to right we plotted the cases $q=0.1$, $q=0.5$ and $q=0.9$.}
    \label{fig:indici}
\end{figure}

\cite{Ernst} discovered that the vacuum stationary axisymmetric solutions to Einstein's field equations can be generated from the equation
\begin{equation}\label{ernst}
(\xi\overline{\xi}-1)\nabla^2\xi=2\overline{\xi}\nabla\xi\cdot\nabla\xi
\end{equation}
with $\xi$ some complex function. In particular, \cite{Ernst} showed that the Kerr metric emerges from the case
\begin{equation}
\xi_E=px-iqy,\quad q=\frac{J}{M^2},\quad p=\sqrt{1-q^2},
\end{equation}
where $x\geq 1$ and $-1\leq y\leq 1$ are prolate spheroidal coordinates, $M$ is the mass of the gravitational source and $J$ its total angular momentum. A generalization of (\ref{ernst}) was found by \cite{tomsato}, namely
\begin{equation}
\xi_T=\frac{p^2 x^4+q^2 y^4-2ipqxy(x^2-y^2)-1}{2px(x^2-1)-2iqy(1-y^2)}.
\end{equation}
Moreover, the relation between the prolate spheroidal coordinates $(x,y)$ and the cylindrical coordinates $(\rho,z)$ are \cite{tomsato}
\begin{equation}\label{polcoord}
\rho=\sigma\sqrt{(x^2-1)(1-y^2)},\quad
z=\sigma xy,\quad
\sigma=\frac{Mp}{\delta}
\end{equation}
with $\delta=1$ and $\delta=2$ for the the Ernst and TS solution, respectively. Finally, note that $\xi_T\to (x^2+1)/2x$ for $q\to 0$ and hence, in this limit the TS solution coincides with the class of Weyl's metrics generated by 
\begin{equation}
\xi_W=\frac{(x+1)^\delta+(x-1)^\delta}{(x+1)^\delta-(x-1)^\delta}
\end{equation}
for $\delta=2$. In the case $\delta=1$, the function $\xi_W$ leads to the Schwarzschild metric \cite{Voor}. Therefore, the parameter $\delta$ can be viewed as a positive parameter measuring the deviation from spherical symmetry. In the following, we consider null-geodesics of the Tomimatsu  solution with $\delta=2$ \cite{tomsato, tomsatop, bosewang, glass, tomsato1, tomsato2,kinkel} in the equatorial plane. These solutions to the Einstein field equations in vacuo are characterised as being stationary, axisymmetric, asymptotically flat and exact. Their generalizations to the cases $\delta=2$ and $3$ are given in \cite{tomsatop}. In geometrized units, the TS space-time is represented by the line element
\begin{equation}\label{tomsato}
ds^2 = fdt^2-2f\omega dtd\varphi-\frac{e^{2\gamma}}{f}\left(d\rho^2+dz^2\right)-\left(\frac{\rho^2}{f}-f\omega^2\right)d\varphi^2,
\end{equation}
where
\begin{eqnarray}
f&=&\frac{A}{B}, \quad \omega =2Mq\left( 1 - y^2\right)\frac{C}{A}, \qquad \text{e}^{2 \gamma} = \frac{A}{p^{4} \left( x^2 - y^2 \right)^{4}},\label{egamma}\\
A&=&[p^2(x^2-1)^2+q^2(1-y^2)^2]^2-4p^2q^2(x^2-1)(1-y^2)(x^2-y^2)^2,\label{Alungo}\\
B&=&(p^2x^4+q^2y^4-1+2px^3-2px)^2+4q^2y^2(px^3-pxy^2+1-y^2)^2,\\
C&=&p^2(x^2-1)[(x^2-1)(1-y^2)-4x^2(x^2-y^2)]\nonumber\\
&&-p^3x(x^2-1)[2(x^4-1)+(x^2+3)(1-y^2)]+q^2(1+px)(1-y^2)^3.
\end{eqnarray}
Using the formulation for non-isotropic spaces introduced in Sec.~\ref{sec:sec2} and in particular, equation  \eqref{nisoindex}, we have the following refractive indices
\begin{equation}\label{rifrazione}
n_\rho=n_z=\frac{e^\gamma}{f}=\frac{B}{p^2(x^2-y^2)^2\sqrt{A}}, \quad n_\varphi=2Mq(1-y^2)\frac{C}{A}+\rho\frac{B}{A}.
\end{equation}
At this point a comment is in order. In the case $q=0$, we find that $A=(x^2-1)^4$ and $B=(x-1)^2(x+1)^4$ and the corresponding refractive indices are
\begin{equation}
n_{\rho,0}=n_{z,0}=\frac{(x+1)^2}{(x^2-y^2)^2},\quad
n_{\varphi,0}=\frac{\rho}{(x-1)^2}.
\end{equation}
Moreover, for $|q|=1$ the TS metric as well as its generalization for different values of the deformation parameter goes over into the line element of an extreme Kerr BH \cite{Bambi}.  In what follows, we suppose that $0\leq q<1$. Then, the refractive indices (\ref{rifrazione}) are real provided that $A>0$. Since this inequality cannot be solved analytically, we plotted in Fig.~\ref{fig:indici} for different values of the parameter $q$ those regions in the $(x,y)$-plane (yellow colour) where the refractive indices (\ref{rifrazione}) are real.  
For the reminder of this section we study the weak lensing problem for null geodesics in the equatorial plane. We start by observing that from the first equation in \eqref{polcoord} we have $\rho\geq 0$ whenever $y^2 \leq 1 \leq x^2$. This implies that $-1 \leq y \leq 1$. Hence, for geodesics on the equatorial plane, i.e. $z = 0$, we must have
\begin{equation} \label{eqplset}
y = 0,\quad 
\rho = \sigma\sqrt{x^2 - 1}
\end{equation}
and the line element (\ref{tomsato}) becomes
\begin{equation}
ds^2_e=f_0 dt^2-2\omega_0 f_0dtd\varphi-\frac{e^{2\gamma_0}}{f_0}d\rho^2-\left(\frac{\rho^2}{f_0}-f_0\omega_0^2\right)d\varphi^2
\end{equation}
with 
\begin{equation} 
f_0=\frac{p^2 x^4-2px^3+2px-1}{p^2 x^4+2px^3-2px-1},\quad
\omega_0=-2Mq\frac{2px^3-px+1}{p^2 x^4-2px^3+2px-1},\quad
\frac{e^{2\gamma_0}}{f_0}=\frac{(p^2x^4+2px^3-2px-1)^2}{p^4x^8}.
\end{equation}
Note that the subindex zero signalizes that the functions in (\ref{egamma}) have been evaluated at $y=0$. The associated Lagrangian is
\begin{equation}
\mathcal{L}=\frac{1}{2}\left(\frac{ds_e}{d\lambda}\right)^2
=\frac{f_0}{2}\dot{t}^2-\omega_0f_0\dot{t}\dot{\varphi}-\frac{e^{2\gamma_0}}{2f_0}\dot{\rho}^2-\frac{1}{2}\left(\frac{\rho^2}{f_0}-f_0\omega_0^2\right)\dot{\varphi}^2
\end{equation}
with $\lambda$ some affine parameter. The stationarity and axisymmetry of the metric lead to the following conserved quantities
\begin{equation}\label{vouk}
f_0\dot{t}-f_0\omega_0\dot{\varphi}=E,\quad
f_0\omega_0\dot{t}+\left(\frac{\rho^2}{f_0}-f_0\omega_0^2\right)\dot{\varphi}=L_z,
\end{equation}
where $E$ is the energy of the massless particle and $L_z$ the projection of its angular momentum along the spin axis of the gravitational object. In the following, we are interested in the case $E>0$ and hence, the distinction between direct ($L_z>0$) and retrograde orbits ($L_z<0$) is controlled by the sign of $L_z$. From (\ref{vouk}), we immediately find that
\begin{equation}\label{cons}
\dot{t}=\frac{1}{f_0}\left[\frac{\omega_0 f_0^2}{\rho^2}L_z+\left(1-\frac{\omega_0^2 f_0^2}{\rho^2}\right)E\right],\quad
\dot{\varphi}=\frac{f_0}{\rho^2}\left(L_z-E\omega_0\right).
\end{equation}
If we impose the null-condition $g_{\mu\nu}\dot{x}^\mu\dot{x}^\nu=0$ on the equatorial plane $z=0$, we end up with the following ODE
\begin{equation}\label{gleichung}
e^{2\gamma_0}\dot{\rho}^2=E^2-\frac{f_0^2}{\rho^2}\left(\omega_0 E-L_z\right)^2.
\end{equation}
The corresponding effective potential is given by
\begin{equation}
U_{eff}=\frac{f_0^2 E^2}{2\rho^2}\left(\omega_0 -b\right)^2.
\end{equation}
where $b=L_z/E$ is the impact parameter. To investigate the presence of circular orbits, we first introduce the rescalings $\widehat{b}=b/M$, $\widehat{\omega}_0=\omega_0/M$, $\widehat{\rho}=\rho/M$ and then, we express the radial variable $\rho$ in terms of $x$. Imposing that the first derivative of the effective potential vanishes leads to the equation
\begin{equation}
f_0\frac{d\widehat{\omega}_0}{dx}+\left(\widehat{\omega}_0-\widehat{b}\right)\left(\frac{df_0}{dx}-\frac{x}{x^2-1}f_0\right)=0,
\end{equation}
which can be cast into the form
\begin{equation}
\frac{N(x,\widehat{b},q)}{D(x,\widehat{b},q)}=0,\quad
N(x,\widehat{b},q)=\sum_{k=0}^9\mathfrak{a}_k x^k,\quad
D(x,\widehat{b},q)=(x^2-1)(p^2 x^4+2px^3-2px-1)
\end{equation}
where $p=\sqrt{1-q^2}$ and 
\begin{eqnarray}
\mathfrak{a}_0&=&-2p(2\widehat{b}-3q),\quad
\mathfrak{a}_1=\widehat{b}-2q,\quad
\mathfrak{a}_2=16p(\widehat{b}-2q),\quad
\mathfrak{a}_3=-4p^2(\widehat{b}+3q),\\
\mathfrak{a}_4&=&-6p(2\widehat{b}p^2-p^2 q+2\widehat{b}-4q),\quad
\mathfrak{a}_5=6p^2(\widehat{b}+q),\quad
\mathfrak{a}_6=4p^3(4\widehat{b}-3q),\quad
\mathfrak{a}_7=-4p^2(\widehat{b}-2q),\\
\mathfrak{a}_8&=&p\mathfrak{a}_7,\quad
\mathfrak{a}_9=\widehat{b}p^4.
\end{eqnarray}
If we look at $q$ as a small parameter, it is not difficult to verify that the roots of the unperturbed polynomial $N(x,\widehat{b},0)$ are $x_0=4$ and $x_\pm=\pm 1$. A straightforward application of perturbation methods for algebraic equations shows that the effective potential admits a critical point $x_c$ which is represented by the following expansion in the perturbation parameter $q$, namely
\begin{equation}
x_c=4-\frac{110 M}{9b_c}q+\left(\frac{3638}{1125}-\frac{220M^2}{9b_c^2}\right)q^2+\mathcal{O}\left(q^3\right),
\end{equation}
where $b_c$ denotes the corresponding value of the impact parameter. In order to classify this critical point, we observe that
\begin{equation}
\left.\frac{d^2 U_{eff}}{dx^2}\right|_{x=x_c}=-\frac{36}{15625}E^2\frac{b_c^2}{M^2}-\frac{688}{46875}E^2\frac{b_c}{M}q+\mathcal{O}\left(q^2\right)
\end{equation}
is negative, and therefore, the equilibrium point $x_{c}$ is unstable. This also signalizes that $x_{c}$ is a maximum for the effective potential. This implies that the associated geodesic is unstable and under any small perturbation, the particle will either fall towards the gravitational source or approach space-like infinity. Note that in the case of vanishing angular momentum $J$, the null circular orbit is located at $\widetilde{x}_c=4$ or equivalently, at $\widetilde{\rho}_c=\sqrt{15}M/2$. In what follows, we are interested in the case of trajectories on the equatorial plane with $b>b_c$. By means of the second equation in (\ref{cons}) and employing the same rescalings introduced above we can rewrite (\ref{gleichung}) as
\begin{equation}\label{egua}
\left(\frac{du}{d\varphi}\right)^2=\frac{(1-u^2)e^{-2\gamma_0(u)}}{4f_0^2(u)\left(\widehat{\omega}_0(u)-\widehat{b}\right)^2}\left[p^2(1-u^2)-4u^2f_0^2(u)\left(\widehat{\omega}_0(u)-\widehat{b}\right)^2\right].
\end{equation}
Let us assume that the observer and the light source are on the equatorial plane. If we observe that $du/d\varphi$ vanishes at the distance of closest approach $x_0=1/u_0$, we end up with the following quadratic equation for $\widehat{b}$
\begin{equation}
p^2(1-u_0^2)-4u_0^2f_0^2(u_0)\left(\widehat{\omega}_0(u_0)-\widehat{b}\right)^2=0,
\end{equation}
whose roots are
\begin{equation}\label{radici}
\widehat{b}_\pm=\widehat{\omega}_0(u_0)\pm\frac{p\sqrt{1-u_0^2}}{2u_0 f_0(u_0)}.
\end{equation}
Note that the term 
\begin{equation}
\frac{1}{\left(\widehat{\omega}_0(u)-\widehat{b}\right)^{2}}=\frac{(p^2-2pu_0+2pu_0^3-u_0^4)^2}{[\widehat{b}+2qu_0(2p-pu_0^2+u_0^3)]^2}
\end{equation}
appearing in (\ref{egua}) when evaluated at $u=u_0$ does not vanish for any value of the rescaled impact parameter and therefore, (\ref{radici}) are the only values of $\widehat{b}$ ensuring that $du/d\varphi=0$ at the distance of closest approach. Taking into account that the TS metric is asymptotically flat, the deflection angle for a light ray trajectory followed along an anticlockwise direction is given by the following formula
\begin{equation}\label{da1}
\Delta\varphi_\pm=-\pi+2\int_0^{u_0}du~\mathfrak{F}_\pm(u,u_0,q),\quad
\mathfrak{F}_\pm(u,u_0,q)=\frac{2f_0(u)e^{\gamma_0(u)}|\widehat{\omega}_0(u)-\widehat{b}_\pm|}{(1-u^2)\sqrt{p^2(1-u^2)-4u^2 f_0^2(u)(\widehat{\omega}_0(u)-\widehat{b}_\pm)^2}}.
\end{equation}
In order to obtain an approximated formula for the deflection angle in the weak lensing regime, we start observing that $x_0\gg 1$ and therefore, $u_0=1/x_0$ can be considered as a small parameter. However, we do not assume that the lens is slowly rotating, i.e. $q\ll 1$. To construct a Taylor expansion in the small parameter $u_0$, we introduce the change of variable $w=u/u_0$ in (\ref{da1}) which we rewrite as follows
\begin{equation}\label{form1}
\Delta\varphi_\pm=-\pi+2\int_0^{1}dw~
\frac{H_1(w,u_0,q)\sqrt{H_2(w,u_0,q}|H_3(w,u_0,q)-H_3(1,u_0,q)\pm H_4(w,u_0,q)|}{\sqrt{p^2(1-u_0^2w^2)-4w^2H_2^2(w,u_0,q)[H_3(w,u_0,q)-H_3(1,u_0,q)\pm H_4(w,u_0,q)]^2}},
\end{equation}
where
\begin{eqnarray}
H_1(w,u_0,q)&=&\frac{p^2-2pu_0 w+2pu_0^3w^3-u_0^4w^4}{p^2(1-u_0^2w^2)},\quad
H_2(w,u_0,q)=\frac{p^2-2pu_0 w+2pu_0^3w^3-u_0^4w^4}{p^2+2pu_0 w-2pu_0^3w^3-u_0^4w^4},\\
H_3(w,u_0,q)&=&\frac{2qu_0^2w(2p-pu_0^2w^2+u_0^3w^3)}{p^2-2pu_0 w+2pu_0^3w^3-u_0^4w^4},\quad
H_4(w,u_0,q)=\frac{p\sqrt{1-u_0^2}}{2H_2(w,u_0,q)}.
\end{eqnarray}
We used Maple to perform the expansion of the integrand in (\ref{form1}) with respect to the small parameter $u_0$ and the subsequent integration. More precisely, we found that at the cubic order in $u_0$, the deflection angle is 
\begin{equation}\label{ablenkung}
\Delta_\pm\varphi=\frac{8}{\sqrt{1-q^2}}u_0+\frac{15\pi-32\pm16q}{1-q^2}u_0^2+\frac{4\Phi_\pm(q)}{3(1-q^2)^{11/2}}u_0^3+\mathcal{O}(u_0^4)
\end{equation}
with
\begin{eqnarray}
\Phi_\pm(q)&=&15q^{10}\pm(60\pi-144)q^9+(253-90\pi)q^8\pm(576-240\pi)q^7+(360\pi-1162)q^6\pm(360\pi-864)q^5+\nonumber\\
&&(1818-540\pi)q^4\pm(576-240\pi)q^3+(360\pi-1237)q^2\pm(60\pi-144)q+313-90\pi.
\end{eqnarray}
In order to get the deflection angle in the small parameter $M/\rho_0$, we recall that $u_0=1/x_0$ where $x_0$ can be expressed in terms of the corresponding value $\rho_0$ by means of (\ref{eqplset}). Hence, we find that
\begin{equation}\label{oben}
u_0=\left(1+\frac{4\rho_0^2}{M^2(1-q^2)}\right)^{-1/2}.
\end{equation}
If we substitute (\ref{oben}) into (\ref{ablenkung}) followed by an expansion in $M/\rho_0$ we obtain
\begin{equation}\label{finale}
\Delta\varphi_\pm=\frac{4M}{\rho_0}+4\left(\frac{15}{16}\pi-2\pm q\right)\frac{M^2}{\rho_0^2}+\left(\frac{155}{3}-15\pi\pm 10\pi q\mp24q+3q^2\right)\frac{M^3}{\rho_0^3}+\mathcal{O}\left(\frac{M^4}{\rho_0^4}\right).
\end{equation}
At this point, some comments are in order. First of all, the above result corrects and improves the weak lensing formula (31) in \cite{bosewang}. More precisely, even though we agree at the first order with \cite{bosewang}, in contrast to the aforementioned result our (\ref{finale}) exhibits an additional factor $15/16$ multiplying $\pi$ which went missing in the expansion performed by \cite{bosewang} to obtain equation (31) therein. Furthermore, our (\ref{finale}) extends (31) at the third order in $M/\rho_0$. Finally, recalling that in the limit $q\to 0$ the TS metric goes over into the Weyl metric with $\delta=2$, the formula (\ref{finale}) can also be used to describe the weak lensing on the equatorial plane of the corresponding Weyl  space-time.

\section{Conclusions and outlook}\label{Con}
In this paper we have explored three space-times: the {\bf{ST}} metric, the Kerr manifold and the {\bf{TS}} geometry with a certain choice for the deformation parameter. We have derived new expressions for the refractive indices of the ST and Kerr metrics. Moreover, we obtained a general formula representing the deflection angle in the weak field approximation valid at first order in the perturbative parameter and for an arbitrary number of spatial dimensions in the case of the ST manifold. In the context of the Kerr metric, we were able to derive a remarkable formula for the deflection angle in the weak lensing regime for the light paths constrained to the equatorial plane of the BH with a low spin value. Regarding the TS metric, we have corrected and improved the accuracy of a weak lensing result previously obtained by \cite{bosewang}. While working on the TS metric we discovered that the related literature is characterized by several mistakes and misprints starting from the derivation of the TS metric and extending until the most recent work on that metric by  \cite{Bambi} where the shadow of a TS gravitational object with $\delta = 2$ is studied. Instead of using the Boyer-Lindquist (BL) coordinates appropriate to the aforementioned case as given by \cite{Yam02}, the author in \cite{Bambi} adopted the BL coordinates for the case $\delta=1$ which corresponds to the Kerr metric. In a forth-coming paper we will focus our attention on the study of circular null orbits, their (in)stability and the so-called Bose conjecture \cite{bosewang} for the TS metric.

\section*{Acknowledgements}
Work by the author PG was supported by the Khalifa University of Science and
Technology under grant number FSU-2021-014. We are grateful to the anonymous referee for his/her comments that helped to improve the present work.

\bigskip

{\bf Data accessibility } This article does not use data.

\end{document}